\documentclass[11pt]{article}
\usepackage{moriond,epsfig}
\usepackage{amsmath,amssymb}

\def\PRL{\em Phys. Rev. Lett.}
\def\PRD{{\em Phys. Rev.} D}

\begin{document}
\vspace*{4cm}
\title{DECAYS OF SUPERNOVA RELIC NEUTRINOS}

\author{ G.L.~FOGLI,$^{1}$ E.~LISI,$^{1}$ 
A.~MIRIZZI,$^1$$^\dagger$~\footnote[0]{$^ \dagger$
({Speaker. E-mail: \tt alessandro.mirizzi@ba.infn.it})
} D.~MONTANINO$^{2}$ }

\address{$^1$ Dipartimento di Fisica and Sezione INFN di Bari,\\ Via Amendola 173,
70126 Bari, Italy \\
$^2$ Dipartimento di Scienza dei Materiali and Sezione INFN di Lecce,\\ Via Arnesano,
73100 Lecce, Italy}
\maketitle\abstracts{
We propose that future observation of supernova relic neutrino (SRN) can be used to
probe neutrino decay models. We focus on invisible (e.g., Majoron) decays, and work out 
the general solution of SRN kinetic equations in the presence of oscillations plus decay.
We then apply the general solution to specific decay scenario, and show that the predicted
SRN event rate can span the whole range below the current experimental bound. Therefore, future
SRN observations will surely have an impact on the neutrino decay parameter space.}

\newcommand{\nuornubar}{{\stackrel{{}_{(-)}}{\nu}\!\!}}
\newcommand{\nubar}{\bar \nu}
\newcommand{\nui}{\nu_i}
\newcommand{\nuj}{\nu_j}
\newcommand{\nubari}{\nubar_i}
\newcommand{\nubarj}{\nubar_j}

\section{Introduction}

In general, massive neutrinos can  not only mix, but  also decay.
The most stringent and safe limit comes from the nonobservation of decay effects on
solar $\nu$ flux. 
However, due to the relatively small distance from the Sun, this limit is very
weak, namely  $ \tau_i / m_i 
\gtrsim 5 \times 10^{-4}
\mathrm{\ s/eV}$,
%..................................................................
  where $\tau_i$ represents the lifetime of the eigenstates $\nu_i$
with mass $m_i$.
 Therefore, the possibility of neutrino decay with longer lifetimes
 (and coming from other astrophysical sources)  cannot be excluded.

Here we focus on the diffuse $\nu$ background  produced by
all  past core-collapse Supernovae (SN) in the Universe---the so-called supernova
relic neutrinos (SRN). 
Future observations of SRN
 can probe decay lifetimes of cosmological interest.
In fact, for SRN decay effects to be
observable in our universe, it must be roughly $\tau_i E/m_i
\lesssim 1/H_0$, where $H_0$ is the Hubble constant. By setting
$H_0=70 \;\mathrm{km}\,\mathrm{s}^{-1}\,\mathrm{Mpc}^{-1}$ and
taking $E\sim O(10)$~MeV (i.e., in the energy range probed by
supernova neutrinos), a rough upper bound for the ``SRN neutrino decay
observability'' is obtained, namely 
%....................................................................
$ {\tau_i}/{m_i}  \lesssim O(10^{11}) \mathrm{\
s/eV} $.
%....................................................................
The comparison of this bound with the previous solar limit
implies that SRN leave  many decades in $\tau_i/m_i$   open to
experimental and theoretical investigations.

Motivated by  this challenging opportunity, in this talk
we aim at showing how to incorporate the effects of both flavor transitions and decays in 
observable SRN spectra, by solving the SRN kinetic equations for two-body non-radiative
$\nu$ decay. This work is based on the results obtained in~\cite{Fogli:2004gy},
 to which we refer
the interested reader for further details.

The plan of the talk is as follows. In Sec.~2 we discuss the general case of $3 \nu$
flavor transitions followed by decays, and give the explicit solution of the neutrino
kinetic equations for generic decay parameters. Specific numerical examples
(inspired by neutrino-Majoron decay models) are given in Sec.~3, in order
to show representative SRN event rates and energy spectra in the presence of decay.
Finally, in Sec.~4 we draw the conclusion of our work.

\section{Three-neutrino flavor transitions and decays}
In this Section we discuss and solve the neutrino kinetic 
equations in the general case of 3$\nu$ flavor transitions
plus decay.
Notice that, for $\tau_i/m_i$ values above the solar $\nu$ bound, 
SRN flavor transitions
occur in matter (and become incoherent) well before neutrino decay
losses become significant, so that hypothetical interference
effects between the two phenomena can be neglected.
For our purposes, flavor transitions inside the supernova can thus
be taken as decoupled from the subsequent (incoherent) propagation
and decay of mass eigenstates in vacuum.
  
\subsection{3$\nu$ flavor transitions}

We assume the active 3$\nu$ oscillation scenario. 
In this framework, the 3$\nu$ squared mass spectrum can be
cast in the form 
$(m^2_1,m^2_2,m^2_3)= M^2+\left(-\frac{\delta
m^2}{2},+\frac{\delta m^2}{2},\pm\Delta m^2 \right)$, 
where $M^2$ fixes the absolute
$\nu$ mass scale; $\delta m^2$ and $\Delta m^2$ govern  two independent
$\nu$ oscillation frequencies, with  
 $|\Delta m^2|\gg \delta m^2$, as indicated by 
current data.
The case of $+\Delta m^2$ ($-\Delta m^2$) characterizes the so-called
 \emph{normal} (\emph{inverted}) hierarchy.
The  elements of the mixing matrix $U$  are parametrized in terms of
three mixing angles $(\theta_{12},\theta_{13},\theta_{23})$.

The yield~\footnote{The yield of a $\nu$
species represents the time-integrated $\nu$ luminosity.}  
 $Y_{\nu_i}$  of the $i$-th mass eigenstate at the surface of the supernova
can be calculated by taking into account the stellar matter effects on 
$\nu$ propagation in SN. These effects are parametrized in term of a level crossing
probability $P_H=P_H(\Delta m^2, \theta_{13})$ 
 among the instantaneous eigenstates of the Hamiltonian (the so
called ``matter eigenstates'') in the dense medium~\footnote{
In principle, there could be an additional level crossing probability
$P_L = P_L(\delta m^2, \theta_{12})$. However, according to current
$\nu$ phenomenology, it is $P_L \simeq 0$ for typical supernovae
density profiles, and so we will neglect it hereafter.} .
The final results
for the yields  
 $Y_{\nu_i}$  at the exit from the
supernova are collected in Table~I in~\cite{Fogli:2004gy}.

\subsection{3$\nu$ decays}

At the exit of SN, $\nu_i$ mass eigenstates evolve indipendently 
until they reach the surface of the Earth. However, 
in their propagation in vacuum, they may decay.
The number density of mass eigenstates $n_{\nu_i}(E,z)$
per
unit of comoving volume and of energy at redshift $z$
 can be
obtained through a direct integration of the neutrino kinetic
equations, as described below.

For ultrarelativistic relic neutrinos $\nu_i$ the kinetic
equations take the form
\begin{equation}
\left[\frac{\partial}{\partial t}-H(t)E
\frac{\partial}{\partial E}-H(t)\right]n_{\nu_i}(E,t)=
R_{\mathrm{SN}}(t)Y_{\nu_i}(E)  + \sum_{m_j>m_i} q_{ji}(E,t) -
\Gamma_i \,\frac{m_i}{E}\,n_{\nu_i}(E,t) \label{Coll}\ .
\end{equation}

The left-hand side (l.h.s.) of Eq.~(1) represents the
Liouville operator for ultra-relativistic $\nu$, where $H(t)$ is the
Hubble constant at the time  $t$. 
The right-hand side (r.h.s.) of Eq.~(1) contains  two
{\em source\/} terms and one {\em sink\/} term. The first source
term quantifies the standard (decay-independent) emission of
$\nu_i$ from core-collapse supernovae, which depends on the supernova
formation rate $R_{SN}(t)$. 
The second source term, in which
%...............................................................................
\begin{equation}
\label{qji} q_{ji}(E,t)=\int_E^\infty dE'
n_{\nu_j}(E',t)\,B(\nu_j\to\nu_i)\Gamma_{j} \,\frac{m_j}{E'}\,\psi_{\nu_j\to\nu_i}(E',E)\
,
\end{equation}
%...............................................................................
quantifies the population increase of $\nu_i$ due to decays from
heavier states $\nu_j$, with decay width $\Gamma_j = 1/\tau_j$,
branching ratio $B(\nu_j\to\nu_i)$ and normalized decay energy spectrum
$\psi_{\nu_j\to\nu_i}$.
The last (sink) term on the r.h.s. of
Eq.~(1) represents the loss of $\nu_i$ due to decay to
lighter states with  width $\Gamma_i$.

Our result is that these equations can be directly integrated, by rewriting
them in terms of the redshift variable $z=z(t)$ and of a rescaled energy parameter
$\varepsilon=\varepsilon(E,z)$.
By replacing back the variable $E=\varepsilon(1+z)$, one obtains
the general solution of the neutrino kinetic equations,
%...............................................................................
\begin{eqnarray}
n_{\nu_i}(E,z) &=& \frac{1}{1+z}\int_{z}^{\infty}
\frac{dz'}{H(z')}\,\left[
R_\mathrm{SN}(z')\,Y_{\nu_i}\left(E\frac{1+z'}{1+z}\right) \right.\nonumber\\
& & +\left.\sum_{m_j>m_i}q_{ji}\left(E\frac{1+z'}{1+z},z'\right)
\right]e^{-m_i\Gamma_i[\xi(z')-\xi(z)](1+z)/E}\ , \label{Solution}
\end{eqnarray}
%...............................................................................
where we have introduced the auxiliary function
\begin{equation}
\label{csi} \xi(z)=\int_0^zdz'\,H^{-1}(z')(1+z')^{-2}\ .
\end{equation}
%...............................................................................
 
In practice, these equations can be integrated numerically by following the
decay sequence, i.e., starting from the heaviest state ($q_{ji}=0$) and ending
at the lightest state ($\Gamma_{i}=0$).

Finally, the $\bar{\nu}_e$ flux at Earth (redshift $z=0$), relevant for the inverse
$\beta$ decay reaction $\bar\nu_e + p \to n + e^+$ in a Cherenkov detector, is given by
\begin{equation}
\label{nubare} n_{\bar\nu_e}(E) \simeq \cos^2\theta_{12}\,
n_{\bar\nu_1}(E,0)+\sin^2\theta_{12}\, n_{\bar\nu_2}(E,0)\,.
\end{equation}

\section{Applications to scenario inspired by Majoron models}

In this section we apply the general results of Sec.~2.2 to some
representative decay scenarios, inspired by Majoron models. 
We consider only nonradiative (invisible) decays of the kind
$\nu_i\to\nuornubar_j+X$,
where an heavier neutrino $\nu_i$ decays into a ligher detectable
(anti)neutrino $\nuornubar_j$ plus an invisible massless (pseudo)scalar
particle $X$, i.e. a ``Majoron''. 

We examine and compare a few representative $3\nu$ decay cases, which
provide SRN yields higher, comparable, or lower than for no decay.
For simplicity, we focus on two phenomenologically interesting cases in which the branching
ratios and decay spectra become model-independent, namely, the
case of quasidegenerate (QD) neutrino masses ($m_i\simeq m_j\gg
m_i-m_j$) and of strongly hierarchical (SH) neutrino masses ($m_i
\gg m_j\simeq 0$). Table~II in~\cite{Fogli:2004gy}
 displays the relevant characteristics
of the QD and SH cases.

%%%%%%%%%%%%%%%%%%%%%%%%%%% FIGURE 1 %%%%%%%%%%%%%%%%%%%%%%%%%%%%%%%%%%%%%%%%%%%%%%
\begin{figure}[!h]
\centering
\vspace*{-5mm}
\hspace*{0mm}\epsfig{figure=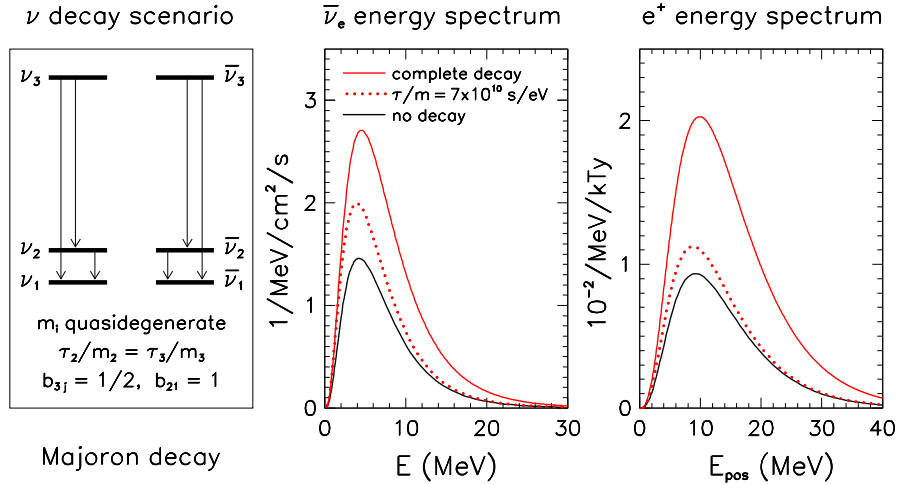, width =0.74\textwidth}
\caption{Supernova relic
$\bar\nu_e$ spectrum (middle panel), and associated positron
spectrum from $\bar\nu_e+p\to n+e^+$ (right panel), for a decay
scenario with normal hierarchy and quasidegenerate masses (left
panel, with $\tau/m$ and branching ratios assignments). 
\label{fig1}}
\end{figure}
%%%%%%%%%%%%%%%%%%%%%%%%%%%%%%%%%%%%%%%%%%%%%%%%%%%%%%%%%%%%%%%%%%%%%%%%%%%%%%%%%%%%%

\subsection{Three-family decays for normal hierarchy and quasidegenerate
masses}

This  decay scenario 
provides SRN densities generally {\em higher\/} than for no decay.
The relevant features of this scenario are graphically shown in 
the left panel of
Figure~1.
The QD approximation
forbids decays of neutrinos into antineutrinos and vice versa
(see Table~II in~\cite{Fogli:2004gy}).

 By construction, the
decay scenario considered in this section is thus governed by just
one free parameter ($\tau/m$). Notice that, for $\tau/m\sim O(10^{10})$~s/eV, SRN
decay effects are expected to occur on a truly cosmological scale.
For much larger values of $\tau/m$,
the no-decay case is recovered. For much smaller values of
$\tau/m$, SRN decay is instead {\em complete\/},  all SRN being in
the lightest mass eigenstate $\bar\nu_1$ at the time of detection.

Figure~1 shows the supernova relic $\bar\nu_e$ energy spectrum,
and the associated (observable) positron spectrum, for the considered decay
scenario. The energy spectra for complete decay (red solid curves)
appear to be a factor of $\sim 2$ higher than for no decay (black solid curve). 

For incomplete neutrino decay (i.e., for $\tau/m\sim
O(10^{10})$~s/eV), one expects an intermediate situation leading
to a SRN flux moderately higher than for no decay. Figure~1
displays the results for a representative case ($\tau/m=7\times
10^{10}$~s/eV, red dotted curves).
In conclusion, the decay scenario examined in this section can lead
to an increase of the SRN rate, as compared with the case of no
decay. The enhancement can be as large as a factor $\sim 2$, the
larger the more complete is the decay.

\subsection{Three-family decays for normal hierarchy and $m_1\simeq 0$}
This  decay scenario 
provides observable SRN densities generally {\em comparable\/} to
the no-decay case. In this case, the
approximation of strong hierarchy (SH) can be applied to the
decays of $\nu_{2,3}$ (and of $\bar\nu_{2,3}$).
Figure~2 shows the supernova relic $\bar\nu_e$ and positron
spectra for this scenario where, as depicted in the left panel,
all decay channels are open. This complex decay chain produces
a substantial
enhancement of the SRN energy spectrum at low energy, visible as a
``pile-up'' of decayed neutrinos with degraded
energy in the middle panel of Fig.~2.
Analogously, the case
of incomplete decay (e.g., $\tau/m=7\times 10^{10}$~s/eV, green
dotted curves), is appreciably different from the cases of no
decay and of complete decay only at low energy.

In this scenario, the interesting effects of decay are almost
completely confined to low $\bar\nu_e$ energies, and are thus
washed out in the observable $e^+$ spectrum, due to the cross
section enhancement of high-energy features. In conclusion, the $e^+$
spectra for the three cases of complete, incomplete, and no decay,
turn out to be very similar to each other (right panel of Fig.~2), and so
in this scanario
 decay effects cannot easily be distinguished  through future SRN observations.

%%%%%%%%%%%%%%%%%%%%%%%%%%% FIGURE 2 %%%%%%%%%%%%%%%%%%%%%%%%%%%%%%%%%%%%%%%%%%%%%%
\begin{figure}[!h]
\centering
\vspace*{0mm}
\hspace*{0mm}
\epsfig{figure=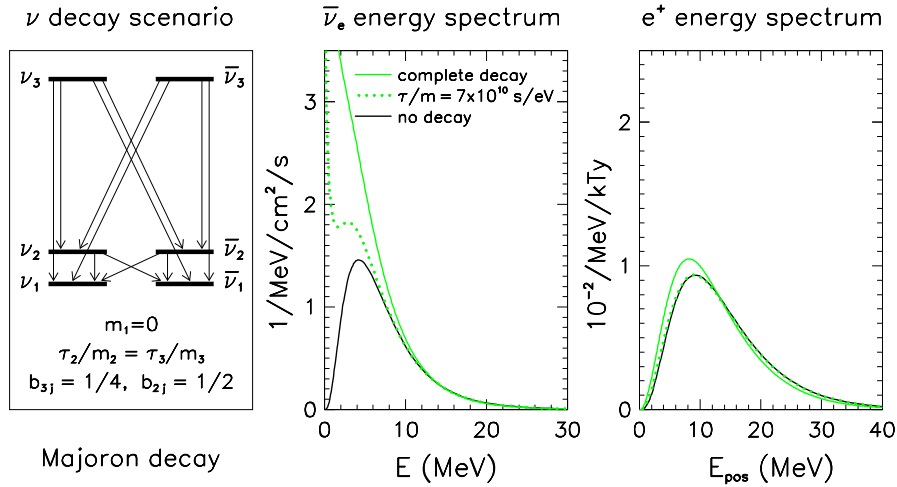, width =0.74\textwidth}
 \caption {\label{fig2}
Supernova relic
$\bar\nu_e$ spectrum (middle panel), and associated positron
spectrum from $\bar\nu_e+p\to n+e^+$ (right panel), for a decay
scenario with normal hierarchy and $m_1\simeq 0$ (left panel, with
$\tau/m$ and branching ratios assignments). 
}
\end{figure}
%%%%%%%%%%%%%%%%%%%%%%%%%%%%%%%%%%%%%%%%%%%%%%%%%%%%%%%%%%%%%%%%%%%%%%%%%%%%%%%%%%%%%

\subsection{Three-family decay for inverted hierarchy}

We conclude our survey of $3\nu$ decays  by discussing a scenario
where the SRN density is generally {\em suppressed\/}, as compared
with the case of no decay. The main features of this scenario are graphically shown 
in the left panel of Figure~3. 
 It results that only the
decay $\bar\nu_2\to\bar\nu_1$ (where the QD approximation is
applicable) is relevant to SRN observations. In fact, decays to
$\bar\nu_3$ provide a negligible amount of $\bar\nu_e$'s ($\propto
\sin^2\theta_{13} \lesssim 10^{-2}$), so that the absolute value of $m_3$ makes no
difference. In the
above scenario, the case of complete decay ($\tau/m\ll
O(10^{10})$~s/eV) is trivial: since the final state is populated
only by $\bar\nu_3$ (and $\nu_3$), the relic density of
$\bar\nu_e$
is negligibly small ($\propto \sin^2\theta_{13}$).
%..................................................................
The nontrivial case of incomplete decay ($\tau/m\sim
O(10^{10})$~s/eV) is then expected to lead to an intermediate
suppression of the SRN density. Figure~3 shows the numerical
results for the specific value $\tau/m=7\times 10^{10}$~s/eV, in
both cases $P_H=1$ (solid curves) and $P_H=0$ (dashed curves). The
neutrino spectra for incomplete decay (blue curves) appear to be
systematically lower than the corresponding no-decay spectra
(black curves), although the difference is mitigated in the
positron spectra (right panel of Fig.~3).

In conclusion, in the inverted
hierarchy scenario of Fig.~3  the SRN signal is generally
suppressed by neutrino decay, and can eventually disappear for
complete decay.

%%%%%%%%%%%%%%%%%%%%%%%%%%% FIGURE 3 %%%%%%%%%%%%%%%%%%%%%%%%%%%%%%%%%%%%%%%%%%%%%%
\begin{figure}[!t]
\centering
\vspace*{-5mm}
\hspace*{0mm}\epsfig{figure=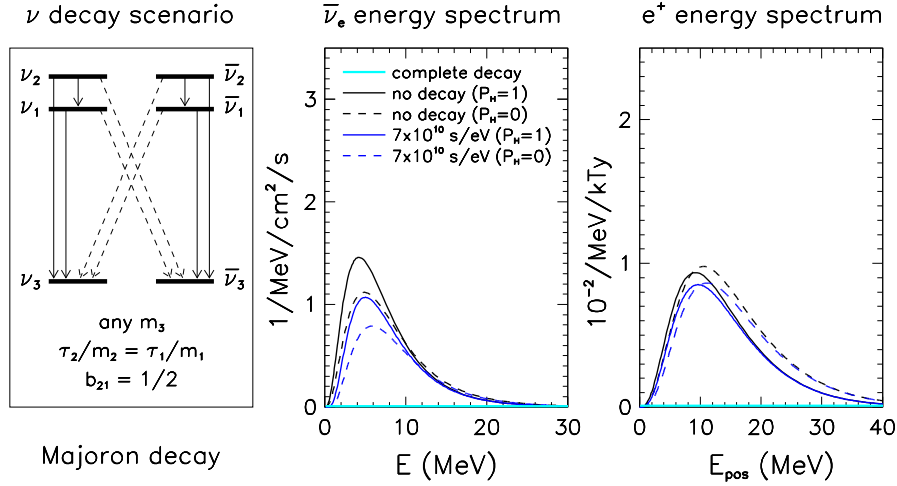, width =0.74\textwidth}
 \caption{\label{fig3} 
 Supernova relic
$\bar\nu_e$ spectrum (middle panel), and associated positron
spectrum from $\bar\nu_e+p\to n+e^+$ (right panel), for a decay
scenario with inverted hierarchy and generic $m_3$ (left panel,
with $\tau/m$ and branching ratios assignments). 
}\end{figure}
%%%%%%%%%%%%%%%%%%%%%%%%%%%%%%%%%%%%%%%%%%%%%%%%%%%%%%%%%%%%%%%%%%%%%%%%%%%%%%%%%%%%%

\subsection{Overview and summary of $3\nu$ decay}

We think it useful to show
also the behavior of the SRN signal for continuous values of the
free parameter $\tau/m$, for the three different decay scenario
discussed above.
Figure~4 shows the positron event rate integrated in the energy
window $E_{e^+}\in [10\, ,\, 20]$ MeV which 
might become accessible to future,
low-background SRN searches~\cite{GADZ}. For each scenario, the
rate is normalized to the standard expectations for no decay and
normal hierarchy (NH), and is plotted as a function of $\tau/m$.
It results that  neutrino decay can enlarge the reference no-decay
predictions for observable positron rates by any factor $f$ in the
range $\sim [0\, ,\, 2.3]$, depending on the particular decay scenario and
provided that $\tau/m$ is in the
cosmologically interesting range below $O(10^{11})$~s/eV.

Since the current experimental upper bound on the SRN flux from SK
 is just a factor of $\sim 2$--3 above typical no-decay
expectations \cite{SKSN},
future observations below such bound are likely to have an impact
on neutrino decay models. If experimental and theoretical
uncertainties can be kept smaller than a factor of two (a
nontrivial task), one should eventually be able to rule out, at
least, either the lowermost or the uppermost values in the range
$f\in [0\,,\,2.3]$, i.e., one of the extreme cases of ``complete
decay.'' Optimistically, one might then try to constrain specific
decay models and lifetime-to-mass ratios through observations.

\begin{figure}[!t]
\centering
\vspace*{7mm}
\hspace*{0mm}\epsfig{figure=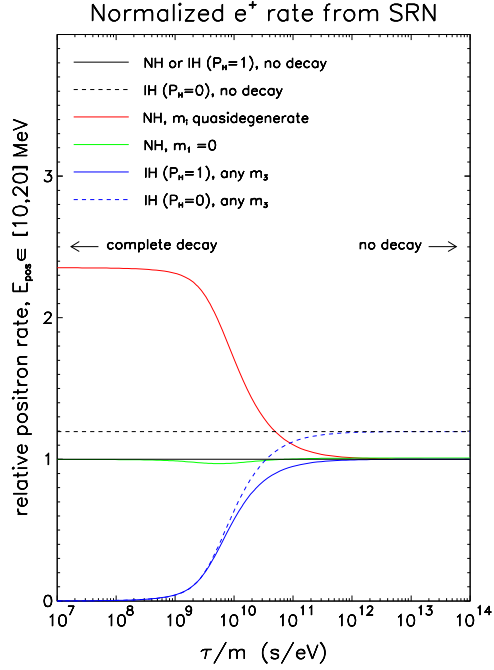, width =0.4\textwidth}
 \caption{\label{fig4}  Positron event
rates in the energy range $[10\, ,\,20]$~MeV for various decay
scenarios, normalized to standard expectations for normal
hierarchy and no decay. Notice how the
expectations branch out (and then reach the complete decay limit)
in the cosmologically relevant range $\tau/m\lesssim
10^{11}$~s/eV.
}
\end{figure}
%%%%%%%%%%%%%%%%%%%%%%%%%%%%%%%%%%%%%%%%%%%%%%%%%%%%%%%%%%%%%%%%%%%%%%%%%%%%%%%%%%%%%

\section{Conclusion}

Neutrino decays with cosmologically relevant neutrino lifetimes
[$\tau_i/m_i \lesssim  O(10^{11})$~s/eV] can, in principle, be
probed through observations of supernova relic $\bar\nu_e$ (SRN).
We have shown how to incorporate the effects of both flavor
transitions and decays in the calculation of the SRN density, by
finding the general solution of the neutrino kinetic equations for
generic two-body nonradiative decays. We have then applied such
solution to three representative decay scenarios which lead to an
observable SRN density larger, comparable, or smaller than for no
decay. In the presence of decay, the expected range of the SRN
rate is significantly enlarged (from zero up to the current upper
bound). Future SRN observations can thus be expected to constrain
at least some extreme decay scenarios and, in general, to test the
likelihood of specific decay models, as compared with the no-decay
case.

\section*{Acknowledgments}
A.M. is sincerly grateful to the organizers of the ``XXXIXth 
\emph{Rencontres de Moriond}''  for 
their kind hospitality in La Thuile.

\end{document}